\documentclass[aps,prd,superscriptaddress,nofootinbib,11pt,tightenlines]{revtex4-1}
\pdfoutput=1

\usepackage[english]{babel}
\usepackage{amsmath,amssymb,amsbsy,amstext,amsthm,booktabs,simplewick,exscale,relsize,slashed,graphicx,amsfonts,upgreek}
\usepackage{tabu,multirow,color,dcolumn,bm,enumerate}
\usepackage{hyperref}
\usepackage{soul}

\newcommand{\gev}{\text{GeV}}

\usepackage{graphicx,bm}

\newcommand{\beq}{\begin{equation}}
\newcommand{\bea}{\begin{eqnarray}}
\newcommand{\eeq}{\end{equation}}
\newcommand{\eea}{\end{eqnarray}}
\newcommand{\bal}{\begin{align}}
\newcommand{\eal}{\end{align}}

\newcommand\brabarb{\scalebox{.3}{(}\raisebox{-1.7pt}[0pt][0pt]{$-$}\scalebox{.3}{)}}

\usepackage{tikz}
\usetikzlibrary{decorations.pathmorphing,decorations.markings}
\tikzset{
photon/.style={decorate, decoration={snake,amplitude=4pt, segment length=7pt}, draw=black},
particle/.style={draw=black, postaction={decorate}, decoration={markings,mark=at position .5 with {\arrow[draw=black]{>}}}},
antiparticle/.style={draw=black, postaction={decorate}, decoration={markings,mark=at position .5 with {\arrow[draw=black]{<}}}},
gluon/.style={decorate, draw=black, decoration={coil,amplitude=3pt, segment length=4pt}},
higgs/.style={draw=black,dashed,thick },
arrow/.style={draw=black, very thick, postaction={decorate}, decoration={markings,mark=at position 1 with {\arrow[draw=black]{>}}}}
}
\begin{document}

\thispagestyle{empty}

\begin{center}

\begin{center}

{\LARGE\sf Reinterpreting $pp \to W^+W^-$ searches for charginos}

\end{center}

\vspace{.5cm}

\textbf{
 Antonio Delgado, Adam Martin
}\\

 {\em Department of Physics, University of Notre Dame, 225 Nieuwland Hall \\ Notre Dame, IN 46556, USA
}

\end{center}

\begin{quote}\small
\begin{center}
\textsc{Abstract:} 
\end{center}
The hallmark way to search for electroweakinos in natural supersymmetry at the LHC involves the trilepton plus missing energy ($\slashed E_T$) final state. This approach assumes an electroweakino mass hierarchy that allows for cascade decays leading to a final state of $W^{\pm}Z^0$ plus $\slashed E_T$. There are, however, situations when that decay pattern may not exist, such as when a chargino is the lightest electroweakino and the lightest supersymmetric particle is the gravitino. In regions of the parameter space where this ordering occurs, the production of any combination of neutralino/chargino leads to a $W^+W^-$+$\slashed E_T\,+ X$ final state, where $X$ could be additional jets or leptons. If $X$ is soft, then all neutralino/chargino production modes fall into the same experimental final state, $\ell^+\ell^- + \slashed E_T$.  ATLAS and CMS have $W^+(\ell^+ \nu)W^-(\ell^-\bar{\nu}) + \slashed E_T$ searches, but their interpretation assumes a spectrum consisting of an isolated charged state. In this paper, we identify the circumstances under which natural supersymmetry models can avoid $W^{\pm}Z^0\,+ \slashed E_T$ bounds. For scenarios that escape $W^{\pm}Z^0\,+\,\slashed E_T$, we then recast the latest ATLAS $W^+W^-+\,\slashed E_T$ search, taking into account all the states that contribute to the same signal. Assuming the lightest supersymmetric particle is massless, we find a bound of 460 GeV for a higgsino-like degenerate doublet. Finally, we extend our arguments to a non-supersymmetric simplified model containing new electroweak-scale $SU(2)_w$ doublets and singlets.

\end{quote}
\vfill
  
\newpage
\section{Introduction}
\label{sec:intro}

 The LHC has several searches for physics beyond the standard model (BSM) that involve large missing transverse energy ($\slashed E_T$). Supersymmetry is one of the prime motivations for these searches, as the lightest supersymmetric particle (LSP) is stable (assuming R-parity), and has to be neutral for cosmological reasons. Motivated by arguments of naturalness and dark matter abundance, we are pushed to a spectrum where electroweakinos (admixtures of electroweak gauginos and Higgsinos) have masses in the range of hundreds of GeV while other supersymmetric particles are much heavier. As electroweakinos cannot decay to squarks/sleptons in this setup, one way to hunt for them is to look in diboson plus missing energy final states. Electroweakinos can decay $\chi \to V+ \text{LSP}$, where $V$ is any electroweak boson (including the Higgs), but decays involving $W^\pm$ and $Z^0$ (typically) dominate in natural setups\footnote{Decays to the Higgs required a large hierarchy of masses among the different neutralinos and decays involving the photon are radiative.}.  Among all possible combinations of electroweakino production and decay mode, $pp\to\chi^0_2\chi^+_1\to Z^0W^\pm \chi^0_1\chi^0_1$ -- associated production of a heavier (non-LSP) neutralino with a chargino followed by their decays to $Z^0/W^\pm$ plus LSP, is the most attractive. Assuming leptonic decays of both the $W^\pm$ and $Z^0$, the final state is very clean, yet it has a largish production cross section (compared to, e.g. $Z^0(\ell^+\ell^-)Z^0(\ell^+\ell^-)$ from a pair of neutralinos) and sufficient handles to suppress the background. Specifically, the SM $W^{\pm}(\ell^{\pm}\nu)Z^0(\ell^+\ell^-)$ background has only one source of missing energy, unlike $W^+(\ell^+\nu)W^-(\ell^-\bar{\nu})$, the background for chargino pair production. In the wide parameter space where this so-called `trilepton' search is applicable, the bound is quite strong, $m_{\chi^{\pm}} \sim m_{\chi_2} \gtrsim 600\, \text{GeV}$~\cite{Sirunyan:2017qaj,Aaboud:2018sua} for a massless LSP.

The prevalence of the trilepton bound in natural supersymmetry leads us an obvious question: what are the circumstances under which $W^{\pm}Z^0 + \slashed E_T$ fails as an electroweakino detection mode? Part of the power of the $W^{\pm}Z^0 + \slashed E_T$ search is its insensitivity to most of the supersymmetry spectrum -- it only cares about the mass hierarchy and decays of the electroweakinos. One easy way to disrupt the $W^{\pm}Z^0 + \slashed E_T$ is to introduce some other BSM state(s) for the electroweakinos to decay to. While an interesting possibility, this necessarily involves adding new light states, taking us away from minimal natural supersymmetry, so we will not pursue this possibility here. The only additional light state we will permit is the gravitino $\tilde G$, a non-electroweakino  neutral state that falls out as the LSP whenever the scale of supersymmetry breaking is low~\cite{Giudice:1998bp}~\footnote{Another possibility leading to the same conclusion would be the siniglino of the NMSSM.}. We will permit additional heavy states, including larger electroweakino sectors, provided all electroweakinos predominantly decay to $W^{\pm}/Z^0 + \text{LSP}$. Second, to narrow our scope further, we will look for scenarios where a different final state, $W^+W^- + \slashed E_T$ takes over as the dominant discovery channel. With these caveats, we can rephrase our focus as: allowing for the possibility of a gravitino LSP, an arbitrary hierarchy of electroweakino soft masses (and $\mu$ term), and additional electroweakinos, what are the criteria for $W^{\pm}Z^{0} + \slashed E_T$ to fail in electroweakino detection while $W^+W^- + \slashed E_T$ succeeds? And in these scenarios, what are the additional experimental consequences?

As we will show, it is possible to avoid $W^{\pm}Z^{0} + \slashed E_T$, even within the MSSM, and in these parameter regions, $pp \to W^+(\ell^+\nu)W^-(\ell^-\bar{\nu}) + \slashed E_T$ is automatically the most sensitive electroweakino detection channel. However, an unavoidable consequence in these scenarios is that the $W^+W^- + \slashed E_T$ final state is populated (at least, at the level of detected, reconstructed objects) by multiple electroweakino production channels ($\chi^+\chi^-, \chi^{\pm}\chi_2$, $\chi_1\chi_2$, etc.). We revisit the $W^+(\ell^+\nu)W^-(\ell^-\bar{\nu}) + \slashed E_T$ results from ATLAS~\cite{ATLAS:2019cfv} and CMS~\cite{Sirunyan:2018lul} to update the bounds including all production channels. In extensions of the MSSM with additional electroweakinos, the realm of possibilities is larger. However, we will show that the dominance of $W^+W^- + \slashed E_T$ over $W^{\pm}Z^0 + \slashed E_T$  is a common outcome in $R$-symmetric extensions of the MSSM~\cite{Kribs:2007ac, Kribs:2008hq}.

Finally, our conclusions regarding the relative sensitivity of $W^{\pm}(\ell^+\nu)Z^0(\ell^+\ell^-)$ and $W^+W^-+\slashed E_T$ (or $\ell^+\ell^- + \slashed E_T$) are not restricted to supersymmetric extensions of the SM. They can be applied to any model with new electroweak multiplets, and we give a simple straw-man example in Sec.~\ref{sec: models}.
 
The structure of the rest of the paper is as follows. In Sec.~\ref{sec: models} we describe the spectra  probed by the different channels and the three different examples where the $W^+W^-+\slashed E_T$ channel  is the most sensitive. In Sec.~\ref{sec:simulation}, we present the details the ATLAS search and  our simulation work. Finally, Sec.~\ref{sec.conclusions} contains our conclusions. Some technical details are presented in the Appendix~\ref{app:mrssm}.

\section{Spectra and Models}
\label{sec: models}

We begin by considering the typical electroweakino spectrum where $W^\pm Z^0 + \slashed E_T$ is applicable. The lightest state, assumed to be neutral for cosmological reasons, sits at the bottom. The next lightest state (NLSP) is neutral, following (increasing in mass) by the first chargino and other neutralinos. We will use LSP for the lightest state throughout this paper, even in circumstances where the LSP is the lightest neutralino ($\chi^0_1$). We will call the next heaviest neutralino $\chi^0_2$, while $\chi^{\pm}$ denotes the lightest chargino. The heavier chargino and heavier neutralinos $\chi^0_3, \chi^0_4$ will play no roll in what follows. A cartoon depicting an example spectrum is shown below in Fig.~\ref{WZ}, where we have taken the LSP to be the lightest neutralino. The fact that the NLSP is a neutralino is important, as its only open decay channel is to the LSP plus something neutral. With all squarks and sleptons decoupled and neglecting loop-level decays to photons, $Z^0 +$ LSP is the only option\footnote{As we are ultimately interested in leptonic final states, we will ignore the possibility of neutralinos decaying to Higgs + LSP.}. The chargino is free to decay to either the NLSP or the LSP by emitting a $W^{\pm}$, though in scenarios where $m_{\chi^{\pm}} \sim m_{\chi_2}$, phase space considerations mean the decay $\chi^{\pm} \to W^{\pm} + $ LSP dominates.

\begin{figure}[h!]
\centering 
\includegraphics[scale=0.8]{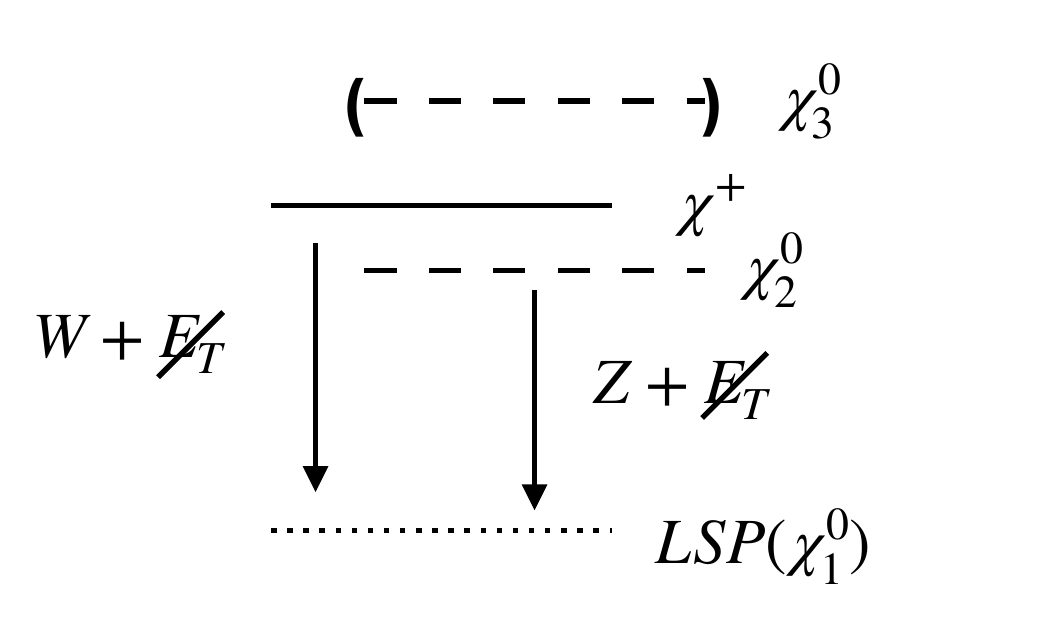}
\caption{Schematic spectrum that will have $W^\pm Z^0$+$\slashed E_T$ as discovery channel. The number of neutralinos depends on the nature (wino or higgino-like) of $\chi_{2,3}^0$.}
\label{WZ}
\end{figure}

In order to suppress $W^{\pm}Z^0 +\slashed E_T$  while keeping $W^+W^-+\slashed E_T$ as a useful search channel for electroweakinos, the simplest possibility is to remove the neutralino -- $\chi^0_2$ in Fig.~\ref{WZ} -- by making it heavy. However, at least within the context of supersymmetry, removing the second neutralino is not feasible. Electroweakinos are part of $SU(2)_w$ multiplets that mix after electroweak symmetry is broken. As a result, the mass difference between the neutral and charged components of the multiplet goes as $\sim m_W^2/M$, where $M$ is the overall mass of the multiplet.  One can not just remove the neutralino from the spectrum without violating $SU(2)_w$ invariance. More generally, in any model where the charged particle belongs to a non-trivial $SU(2)_w$ multiplet, electroweak invariance prevents splitting the different components by arbitrarily large values -- and any term that distinguishes among the different members of the multiplet have to be proportional to the vacuum expectation value of the Higgs. 

However, as will review below, it is possible to reverse the mass order of the lightest chargino and the lightest neutralino. With no further additions, this mass ordering is not acceptable cosmologically, as the lightest electroweakino is now charged, but if we embed the scenario in the context of gauge mediated supersymmetry breaking (or another setup with supersymmetry breaking at a low scale) then the gravitino is the LSP and there is no immediate issue\footnote{The gravitino as DM candidate has some challenges from the model builidng point of view~\cite{Steffen:2006hw} but any discussion in this direction is beyond the scope of this paper since the solutions do not involve the electroweakino spectrum.}. 

With the gravitino as LSP, an NLSP chargino will decay to $W^{\pm} + $ LSP.  The lightest neutralino can decay directly to $Z^0 + $ LSP, however it now also has the possibility to decay to $W^{\mp} + \chi^{\pm}$ (beta decay). If the latter dominates sufficiently, $Z^0$ decays are eliminated and the trilepton signal is stifled.
 
Finally, even if beta decay of neutralinos dominates, neutralino production followed by beta decay and $\chi^{\pm} \to W^\pm + \text{LSP}$ can still lead to a trilepton signal, i.e. $pp \to \chi_2 \chi^{\pm} \to W^\pm \chi^+\chi^- \to 3 \ell + $$\slashed E_T$.  This contribution can be suppressed as well if $\chi^0_2$ and $\chi^{\pm}$ have similar mass. In this case, the lepton from the beta decay is too soft to pass the detector id requirements. 

\begin{figure}[h!]
\centering 
\includegraphics[scale=0.8]{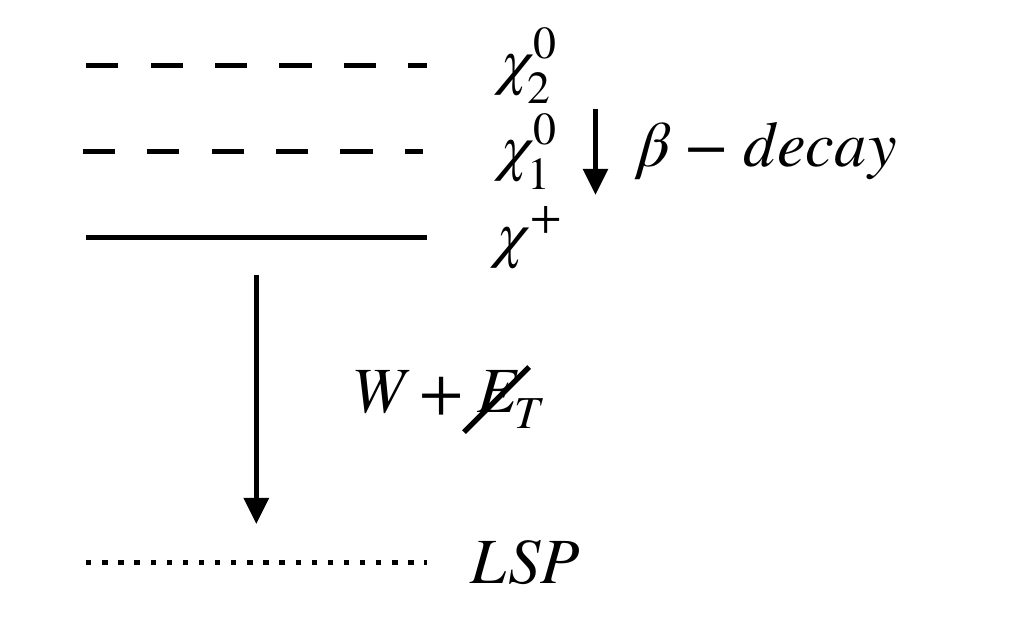}
\caption{Schematic spectrum that will have $W^+W^-$+$\slashed E_T$ as discovery channel. The mass splitting between the neutralinos and the chargino is small so the products from $\beta$-decay are very soft. The number of neutralinos is model dependent but it does not affect the conclusion. In this case, the LSP is the gravitino.}
\label{spectrum}
\end{figure}

While this set of requirements removes (or at least strongly suppresses) the trilepton signal, there are now several channels  ($\chi^+\chi^-, \chi^{\pm}\chi_2$, $\chi_1\chi_2$, etc.) contributing to the $\ell^+\ell^-\,+\,\slashed E_T$ final state and must be considered when interpreting the $W^+W^-$ channel bound. In Fig.~\ref{spectrum} we summarize the necessary criteria for $W^+W^-$+$\slashed E_T$ to be the most sensitive channel.
  
To better illustrate how these requirements work and what they demand of the spectrum, we now introduce three benchmark scenarios. In the context of supersymmetry, the electroweakino sector has been vastly studied in the literature. In particular, Ref.~\cite{Kribs:2008hq} precisely analyzed the conditions for the chargino to be the lightest of the electroweakinos in two different supersymmetric models. We now proceed to summarize what was found in Ref.~\cite{Kribs:2008hq} and to quote the results relevant to our paper.

Starting within the MSSM, and in the limit when one assumes the electroweak breaking  effects are small, the masses of neutralinos can be written as~\cite{Martin:1997ns}:
\begin{eqnarray}
m_{\chi^0_B} &=&M_1-\frac{m^2_Zs^2_W(M_1+\mu\sin 2\beta)}{\mu^2-M^2_1}\nonumber\\
m_{\chi^0_W} &=&M_2-\frac{m^2_W (M_2+\mu\sin 2\beta)}{\mu^2-M^2_2}\nonumber\\
m_{\chi^0_{H1}} &=&|\mu|+\frac{m_Z^2(I-\sin 2\beta)(\mu+M_1 c^2_W+M_2 s^2_W)}{2(\mu+M_1)(\mu+M_2)}\nonumber\\
m_{\chi^0_{H2}} &=&|\mu|+\frac{m_Z^2(I+\sin 2\beta)(\mu-M_1 c^2_W-M_2 s^2_W)}{2(\mu-M_1)(\mu-M_2)}
\label{neut}
\end{eqnarray}
\noindent
where $M_1$ and $M_2$ are assumed to be real and positive and $I$ is equal to $\pm 1$ depending on the sign of $\mu$. Here we are using $B,W,H$ to label different eigenstates according to their nature (mostly bino, wino or higgsino), and $1,2,3,4$ to label their mass from light to heavy. There are similar expressions for the chargino masses:

\begin{eqnarray}
m_{\chi^{\pm}_W} &=&M_2-\frac{m^2_W (M_2+\mu\sin 2\beta)}{\mu^2-M^2_2}\nonumber\\
m_{\chi^{\pm}_H} &=&|\mu|+\frac{I m^2_W(\mu+M_2 \sin 2\beta)}{\mu^2-M_2^2}
\label{charg}
\end{eqnarray}

If we assume that the lightest chargino is wino-like, then $M_2 \ll |\mu|$ so that $m_{\chi^{\pm}_W}$ corresponds to $m_{\chi_1^+} $. In this limit, we see that $m_{\chi^0_W}$ is actually equal to $m_{\chi^{\pm}_1}$ (they will be split by QED corrections), and thus production of the lightest chargino in association with  this wino-like neutralino is roughly the same size (at least from a kinematics perspective) as chargino pair production. If we further take $M_1 \ll M_2, |\mu|$ such that the bino is the LSP, charginos will decay $\chi^{\pm} \to W^{\pm} +\,\text{LSP}$, neutralinos as $\chi^0_2 \to Z^0+\,\text{LSP}$, exactly the type of scenario where (for all values of $M_1, M_2$) $W^\pm Z^0$+$\slashed E_T$ searches are relevant. The same situation will happen if $M_1$ is larger than $M_2$ but another neutral state like the gravitino in GMSB or the singlino in the NMSSM is the LSP.

If we instead assume that higgsinos are lighter than the wino, corresponding to $|\mu| \ll M_2$, then the exact ordering among the charginos and neutralinos depends on the value of $M_1$. If $M_1$ is smaller than $|\mu|$ and $M_2$, then $m_{\chi^0_B}$ corresponds to the LSP and one can easily see that $m_{\chi^0_{H1}}$ and $m_{\chi^0_{H2}}$ are of the same size as $m_{\chi^{\pm}_H}$, and that one of the two is lighter than $m_{\chi^{\pm}_H}$ and the other heavier. The spectrum again matches Fig.~\ref{spectrum},  so $W^\pm Z^0$+$\slashed E_T$ will set the strongest bound.

However, if $M_1 > M_2 > |\mu|$  with positive $\mu$ and $\tan\beta$ close to one, we find the following mass hierarchy for the lightest electroweakinos:
\begin{eqnarray}
m_{\chi^{\pm}} &=& \mu-\frac{m^2_W}{M_2}\nonumber\\
m_{\chi^0_1} &=& \mu-\frac{m^2_W}{2M_2}\nonumber\\
m_{\chi^0_2} &=&\mu
\end{eqnarray}
\noindent
with the other electroweakinos heavier. In this situation, the lightest of the three eigenstates is the charged one. Note that, although one gets a spectrum where the chargino is lighter than any neutralino, the mass difference between $\chi^{\pm}$ and $\chi^0_{1,2}$ is small.  The LSP in this case has to be another neutral state like the gravitino (in GMSB) or the singlino in the NMSSM. We are going to focus in the first possibility as an example where the LSP is not an electroweakino. The NMSSM adds an extra singlet superfield and modifies the neutralino max matrix adding more parameters. Nonetheless, in situations where the LSP is mostly singlino and a chargino is the NLSP, one can realize a spectrum leading to the same signal~\cite{Wang:2019biy}.

Focusing on this mass ordering within the context of GMSB, the $\chi^0_{1,2}$ neutralinos can either beta decay to $\chi^{\pm}$ or directly to the gravitino emitting a photon or a $Z^0$. As explained earlier, for $W^\pm Z^0 + \slashed E_T$ to be suppressed, beta decay of the neutralino must dominate. Ref.~\cite{Kribs:2008hq} introduced the following ratio to distinguish which decay dominates:

\begin{equation}
R_{\Gamma}\equiv \dfrac{ \Gamma(\chi^0_{1,2}\rightarrow \chi^{\pm}\bar{f}f') }{ \Gamma(\chi^0_{1,2} \rightarrow \widetilde{G}X) }~, \label{eq:Rgamma}
\end{equation}
 
The decay of the neutralino to the gravitino is given by~\cite{Giudice:1998bp}:
 
 \begin{equation}
 \Gamma({\chi}^{0}_{1,2}\to \widetilde{G}X) =\kappa_{\widetilde{G}X}\dfrac{m_{\chi^{0}}^{5}}{96\pi M_{P}^{*2}m_{3/2}^{2}}\left[ 1-\dfrac{M_{X}^{2}}{m_{{\chi^{0}}}^{2}} \right]^{4},
 \label{gravitinodecay}
\end{equation}

\noindent
where $\kappa_{\widetilde{G}X}$ encodes the $O(1)$ coupling of the neutralino to the gravitino and $X$ (which introduce a small model dependence into $R_\Gamma$) and $m_{3/2}$ is the gravitino mass. The three-body decay of the neutralino to the chargino and soft leptons is beta decay which makes $R_\Gamma$ proportional to $m_{3/2}^{2}(\Delta m/m_{\chi^\pm})^{5}$, where  $\Delta m$ is the mass difference between the (lightest) neutralino and the (NLSP) chargino. 

The condition for  $W^+W^-$+$\slashed E_T$ to give a stronger bound than  $W^\pm Z^0 + \slashed E_T$ is:

\begin{equation}
\sigma(pp\to\chi^+\chi^-)+\sigma(pp\to\chi^{\pm}\chi^0_{1,2})BR(\chi^0_{1,2}\rightarrow \chi^{\pm}\bar{f}f')>\sigma(pp\to\chi^\pm\chi^0_{1,2}) BR(\chi^0_{1,2} \rightarrow \widetilde{G}X)
\label{ww}
\end{equation}

\noindent
where we are assuming both channels to have similar analysis efficiencies and we are neglecting terms that go quadratically with the branching ratios.  The inequality~\ref{ww} is satisfied when  $R_\Gamma>\frac{\sigma(pp\to\chi^\pm\chi^0_{1,2})-\sigma(pp\to\chi^+\chi^-)}{\sigma(pp\to\chi^\pm\chi^0_{1,2})+\sigma(pp\to\chi^+\chi^-) }$, which is roughly equal to $1/2$  for the values of masses that we are considering. This is a rather simple assumption that does not take into account the specifics of the different final states, such as efficiencies or SM backgrounds, but the important point is that there will always be a region where $W^+W^-$+$\slashed E_T$ will dominate.

\begin{figure}[h!]
\centering 
\includegraphics[scale=0.7]{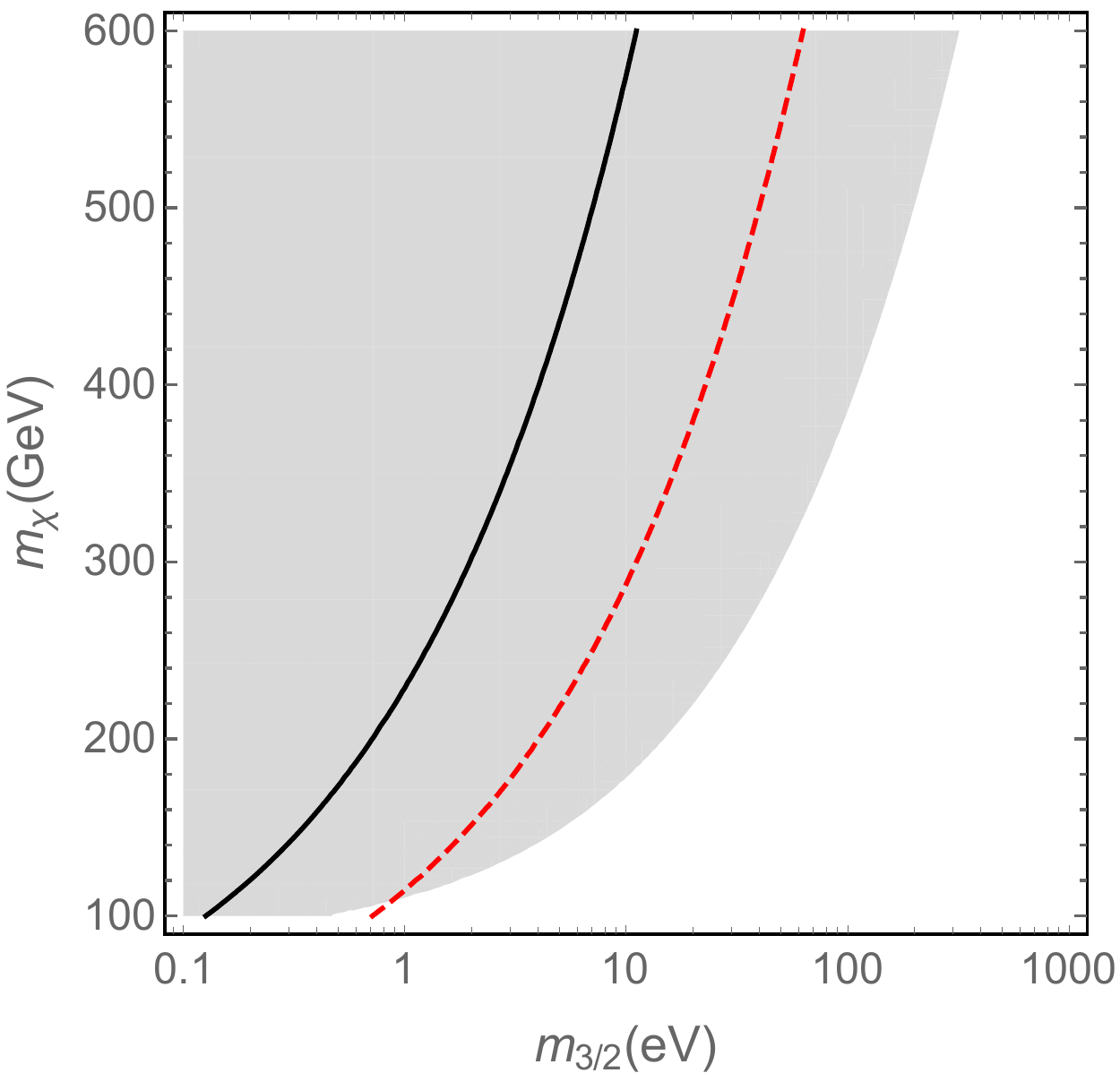}
\caption{Line of $R_\Gamma=1/2$ for $\Delta m=5$ GeV (dashed) and $\Delta m=10$ GeV (solid) in the chargino mass versus gravitino mass plane. The shaded region corresponds to a prompt decay of the chargino. The region right of the lines while in the shaded region is the parameter space where $W^+W^-$+$\slashed E_T$ will give the strongest bound in this model.}
\label{Rgamma}
\end{figure}

In Fig.~\ref{Rgamma}, we have plotted the line of $R_\Gamma  =1/2$ (larger values for $R_\Gamma$ will be to the right of the one plotted) in the plane $(m_{3/2}, m_\chi)$ for $\Delta m=5$ GeV (dashed) and 10 GeV(solid). As can be seen from the figure, the relevant parameter space corresponds to chargino masses of few hundreds of GeV and gravitinos in the tens of eV. The smaller we take $\Delta m$, the smaller $m_{3/2}$ must be to maintain $R_\Gamma > 1/2$. The gravitino mass sets the overall chargino lifetime, and for sufficiently small $m_{3/2}$ the charginos become long-lived. As our focus is on the $W^+(\ell^+\nu)W^-(\ell^-\bar{\nu}) + \slashed E_T$ search, which assumes a promptly decaying signal, we will not consider long-lived electroweakinos in this paper. Long lived electroweakinos are an interesting possibility, but require completely different search strategies; see~\cite{LL1, LL2, Alvarado:2018rfl}. To fix the parameter space, we will designate decay lengths smaller than $0.5$ mm as prompt, corresponds to the shaded region of Fig.~\ref{Rgamma}.

Since the mass splitting $\Delta m$ controls the energy of remnants from $\chi^0_2$ beta decay (which may include additional leptons), if we try to push the parameter space to larger $\Delta m$, we cannot ignore the $W^{\pm}Z^0 + \slashed E_T$ channel -- the lepton $p_T$ requirement in the ATLAS trilepton searches~\cite{Aaboud:2018sua} is set to $10$ GeV. While the turn on of the $W^{\pm}Z^0 + \slashed E_T$ sensitivity will not be immediate at $\Delta m = 10\,\gev$,  we will focus on $\Delta m$ between 5 (smaller values will lead to a long lived chargino for the values of $m_{3/2}$ we are considering) and 10 GeV. Since $m_{3/2} < 100$ eV when translating experimental constraints into this scenario, we must be careful to use interpretations that also assume a massless LSP.

One additional feature of this scenario that is worth mentioning is that the neutralinos are predominantly higgsino like and are therefore pseudo-Dirac~\cite{Choi:2008pi}. As a result, the production of same-sign charginos -- coming from neutralino decays and leading to a final state of same sign leptons plus $
\slashed E_T$ (i.e., $ p p\to \chi_1 \chi_2 \to \chi^+ \chi^+ \tilde G \tilde G\to W^+(\ell^+ \nu)W^+(\ell^+\nu) +\slashed E_T$) -- is very suppressed.  The origin of the suppression is an approximate charge conjugation symmetry only broken by the small splitting between the two neutralinos~\cite{Choi:2008pi}.

While it is possible to arrange for $m_{\chi^{\pm}} < m_{\chi^0_1}$ in the MSSM, the parameter space is quite limited. However, in extensions of the MSSM with Dirac gaugino masses, $m_{\chi^{\pm}} < m_{\chi^0_1}$ is far more common. Some scenarios that contain Dirac gauginos include extra-dimensional supersymmetry models~\cite{Barbieri:2000vh, XD1, XD2, XD3} or 4D models where the $U(1)_R$ symmetry present in the supersymmetric kinetic term is imposed on the rest of the theory~\cite{Fox:2002bu, Dirac1, Kribs:2007ac, Dirac2, Dirac3}. In these so-called R-symmetric models,  there are actually four charginos: the two states from the MSSM, one from the $SU(2)_w$ adjoint Dirac partner of the wino, and one from an additional $SU(2)_w$ doublet (an $R$-Higgs) whose presence is required to impose exact $R$ symmetry on the Higgs terms in the superpotential. For reference, the superpotential for this setup is shown in Appendix~\ref{app:mrssm}. These four states can be further classified by their $R$-charge $(\pm 1)$, so the chargino mass matrix splits into two $2 \times 2$ blocks. In the limit of large $\tan\beta$ and a vanishing $SU(2)_w$ adjoint vev, the $2 \times 2$ block containing the lightest eigenvalue simplifies to 
\begin{equation}
M_{\chi^{\pm}} =  \begin{pmatrix}
 M_{D2} & O(g v/\sqrt{2}) \\
O(\lambda v/\sqrt{2}) & \mu \\
 \end{pmatrix}
 \label{eq:RCmatrix}
 \end{equation}
 \noindent
 where $M_{D2}$ is the $SU(2)_w$ gaugino Dirac mass, $g$ is the $SU(2)_w$ coupling and $\lambda$ is the coupling for trilinear superpotential interaction containing the $SU(2)_w$ adjoint, a MSSM Higgs and its $R$-Higgs partner (see Appendix~\ref{app:mrssm}).

The neutralinos in this setup form four Dirac fermions\footnote{The wino, bino and two Higgsinos of the MSSM (4 Weyl fermions) plus their $R$-symmetric partners (4 more Weyl states)}. In the large $\tan\beta$, small $SU(2)_w$ adjoint vev approximation used above\footnote{In addition, we have set the two superpotential Higgs masses to be the same, $\mu_u = \mu_d = \mu$. See the Appendix for more details}, their mass matrix has the form:
 \begin{equation}
 M_{\chi^0} = 
\begin{pmatrix}
M_{D1} & 0 & 0 & O (g v/2) \\
0 & M_{D2} & 0 &  O (gv/2)\\
0&  0 & \mu  & 0 \\
O(\lambda v/\sqrt{2}) & O(\lambda v/2) & 0 & \mu 
 \end{pmatrix},
 \label{eq:RNmatrix}
 \end{equation}
where the new parameter, $M_{D1}$ is the $U(1)_Y$ Dirac gaugino mass. Let us decouple $M_{D1}$, analogous to what we did in the MSSM case, reducing the neutralino mass matrix to $3 \times 3$. Comparing the lower right $3 \times 3$ block of Eq.~\eqref{eq:RNmatrix} with Eq.~\eqref{eq:RCmatrix}, we see the mass matrices have the same structure, but that the off-diagonal entries are larger for the charginos by a factor of $\sqrt 2$. Diagonalizing, the larger off-diagonal entries translate to larger splitting among eigenvalues, and thus the lightest eigenvalue of $M_{\chi^{\pm}}$ will be lighter than the lightest of $ M_{\chi^0} $. This result is not restricted to the simplifying limits we have taken here and persists throughout large regions of parameter space, as explored thoroughly in Ref.~\cite{Kribs:2008hq}.

 As in the MSSM case, $m_{\chi^{\pm}} < m_{\chi^0_{1,2}}$ is only viable cosmologically in the context of low-energy supersymmetry, where the chargino plays the role of the lightest of the electroweakinos and the gravitino is the LSP. While the R-symmetric model has more parameters and states, we have not introduced any additional light particles (compared to the MSSM in GMSB), so the relevant parameters and kinematics of electroweakino decay is unchanged from our MSSM discussion. As such, Eq.~\eqref{eq:Rgamma} continues to hold in the R-symmetric case, with the viable parameter space described by Fig.~\ref{Rgamma}. Of course, while the mass parameters ($\Delta m, m_{\chi}$, etc.)  are the same, their description in terms of UV parameters is different in R-symmetric models than in the MSSM. Finally, as all electroweakinos are Dirac in this model, there is no possibility for a same sign dilepton signal because there is an exact charge conjugation symmetry~\cite{Choi:2008pi}.
 
% \adam{useless statement, to be repeated later. What we need is an actual transition} We are going to build an ad-hoc model that will incorporate the necessary ingredients for   $W^+W^-$+ $\slashed E_T$ to give the strongest bound.  
 
 While supersymmetry was useful for providing context for the previous two scenarios, our arguments regarding the validity of $W^+W^-$+ $\slashed E_T$ vs. $W^\pm Z^0$+$\slashed E_T$ are essentially just statements about the mass ordering of states in new $SU(2)_w$ multiplets and do not require full supersymmety structure. To illustrate this, consider a simple SM extension consisting of a vector-like fermion $SU(2)_w$ doublet with $Y=-1/2$, $\Sigma$, and a neutral pseudoscalar $\phi$. Adding interactions among $\Sigma$, $\phi$ and the SM, we have: 
\begin{equation}
\mathcal{L}=M_{\Sigma}\overline{\Sigma}{\Sigma}+ \frac{m^2}{2} \phi^2+ \frac{c_\phi}{\Lambda}\partial^\mu \phi\overline{\Sigma}\gamma_\mu\gamma_5 L+\frac{c_h}{\Lambda} (H\Sigma)^2+\dots
\label{eff}
\end{equation} 
\noindent 
 where $L$ is a SM lepton doublet (either $e$ or $\mu$), and for simplicity we have suppressed a possible interaction between $\overline{\Sigma}$, $E$ and the Higgs. This can be justified either by taking the coupling to be very small, or by imposing a discrete symmetry that distinguishes $\phi$ and the Higgs\footnote{For example,$Z_2$ symmetry under which $\Sigma$ and $\phi$ are odd and the rest of the fields even.}. The mass splitting between the states in $\Sigma$ is controlled by the last term and is $\sim v^2/\Lambda$.  For the appropriate sign of $c_h$, the neutral state ($\Sigma^0$) will be heavier than the charged one ($\Sigma^{\pm}$), and it can either decay $\Sigma^0 \to \phi\, \nu$ via the non-renormalizable operator or it can beta decay to the charged component, $\Sigma^0 \to \Sigma^{\pm} \ell^{\mp}\overset{\brabarb}{\nu}$. If $\Sigma^0 \to \phi\, \nu$ dominates, then $\Sigma^0$ production will lead to a purely invisible final state. However, if beta decay dominates, then we will be in a situation, as earlier, where the whole $\Sigma$ doublet contributes to the $\ell^+\ell^- + \slashed E_T$ final state and can be picked up by ATLAS/CMS $W^+W^- + \slashed E_T$ searches\footnote{$\Sigma$ production does not generate any on-shell $W^\pm$'s, but this is irrelevant as the analysis only looks for sufficiently energetic, opposite sign leptons and missing energy.}. By construction, there is no possibility of a same sign dilepton signal in this model.  

\begin{figure}[h!]
\centering 
\includegraphics[scale=0.7]{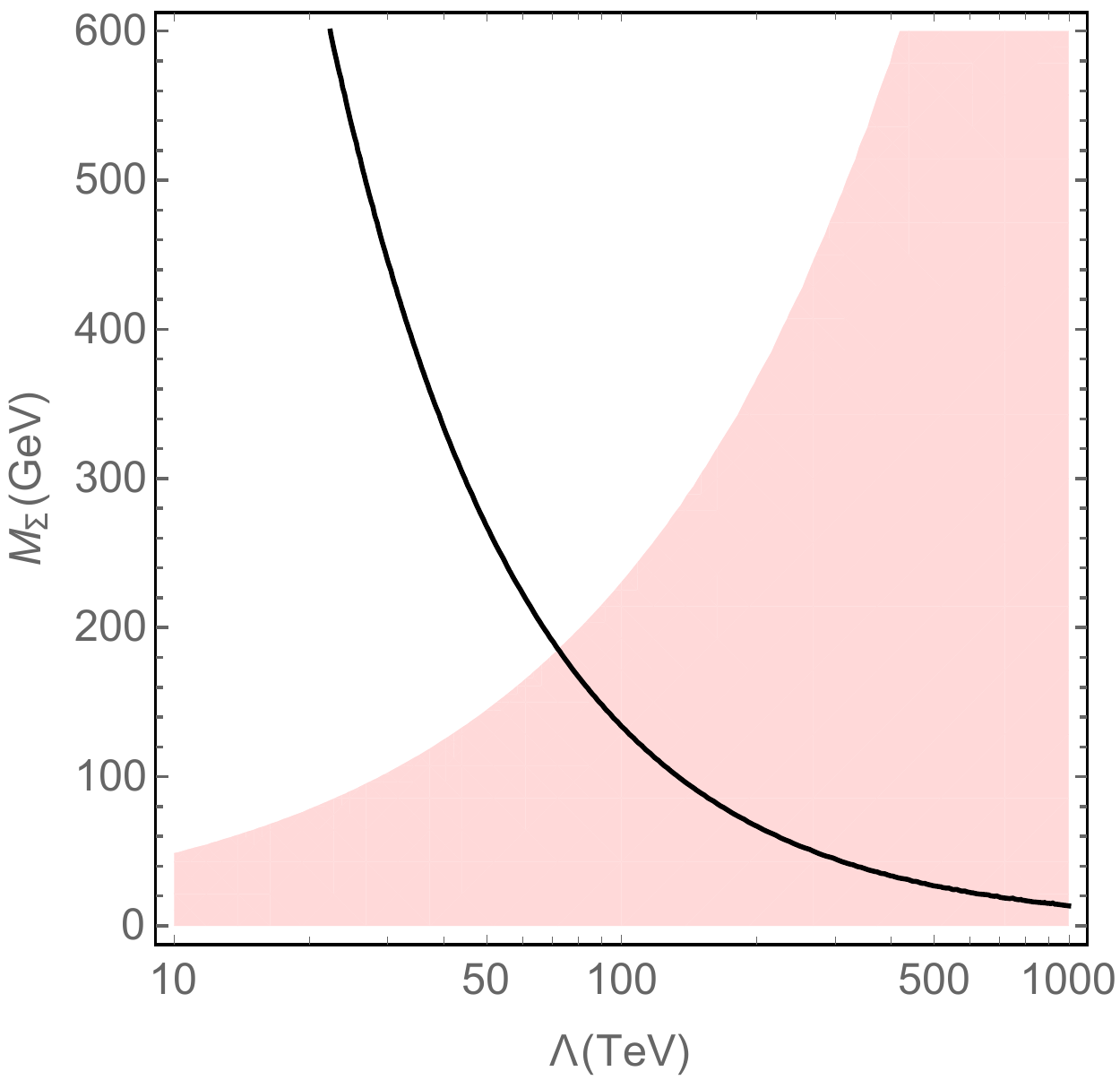}
\caption{The black line indicates where $\Gamma_\beta=\Gamma_\phi$ for $c_h=1$ and $c_\phi=10^{-4}$ as a function of the mass of $\Sigma$ ($M_{\Sigma}$) and $\Lambda$. The white region corresponds to a prompt decay of the charged component whereas the shaded region corresponds to a long-lived charged particle. The viable region is the white one to the left of the black line.} 
\label{Eff}
\end{figure}

Taking $\phi$ to be massless, the partial decay width of $\Sigma^0\to \phi\, \nu$ (denoted as $\Gamma_\phi$) and the partial width of $\Sigma^0\to \Sigma^+ f f'$ (denoted as $\Gamma_\beta$) are equal to:
\begin{align}
\Gamma_\phi&=\frac{c_\phi^2}{8\pi}\frac{M^3_{\Sigma}}{\Lambda^2}\nonumber\\
\Gamma_\beta&=\frac{2}{15\pi^3}\frac{c_h^5}{\Lambda^5} v^6
\end{align}
\noindent
where $v$ is the vev of the Higgs. We can now investigate the region of the parameter space where beta decay dominates. In Fig.~\ref{Eff}, we show the line $\Gamma_\beta=\Gamma_\phi$ in the $\Lambda$ versus $M_{\Sigma}$ plane for $c_h=1$ and $c_\phi=10^{-4}$. To the left of the black line is the viable region, where $\Gamma_\beta>\Gamma_\phi$. The white region corresponds to the prompt decay of $\Sigma^{\pm}$. The splitting between the neutral and the charged component of $\Sigma$ is less than few GeV in the whole plot (and we are still taking the mass of $\phi$ to be negligible to make sure all decays are prompt).
Varying $c_\phi$ and $c_h$ will just move the position of the black line and the region where the decay is prompt but the conclusion will be the same -- there are regions where the whole doublet contributes to the $pp \to \ell^+\ell^- + \slashed E_T$ final state.

Amusingly, our experience with this toy model allows us to craft a scenario with an isolated charged state -- the type of signal the ATLAS/CMS analysis assumes. Instead of the $SU(2)_w$ doublet $\Sigma$, we could introduce a vector like singlet with $Y=-1$ ($\Delta$)  that decays to the right handed leptons plus $\phi$ via the following higher dimensional operator $ \partial^\mu \phi\overline{\Delta}\gamma_\mu\gamma_5 e_R$.

Summarizing this section, we have shown that in scenarios where $W^\pm Z^0$+$\slashed E_T$ will not give any bound due to the suppression of decays $Z^0$'s  the $W^+W^-$+$\slashed E_T$ search will set the strongest bounds. We have also shown that, in these scenarios, there are several states contributing to the same final state and therefore the experimental bound calculation is more complex than the naive experimental interpretation. For example, in the MSSM  suppressing the  $W^\pm Z^0$+$\slashed E_T$ implies that the whole Higgsino doublet will be the lightest of the electroweakinos and will have neutral states contributing to the same final state (plus soft objects) in the decays to the gravitino LSP. The question that remains is how those extra states affect and modify the ATLAS/CMS interpretation.
  
 \section{WW+$\slashed E_T$ search and Simulation}
\label{sec:simulation}

Both ATLAS~\cite{ATLAS:2019cfv} and CMS~\cite{Sirunyan:2018lul} have searches for two opposite sign (OS) leptons and missing energy that they interpret as a bound on chargino pair production, where the charginos decay to a leptonic $W^{\pm}$ and a neutralino LSP ($pp\to\chi^+\chi^-\to W^+(\ell^+\nu)W^-(\ell^-\nu)\chi_1^0\chi^0_1$).  The CMS search uses 36 fb$^{-1}$ and excludes chargino masses between 160 and 200 GeV for a massless LSP, while the ATLAS variation uses the full Run II luminosity at 13 TeV, 138 fb$^{-1}$.  For a purely wino-like (i.e. charged components of an $SU(2)$ triplet, with the other components are taken to have no role in the bound) and decaying $100\%$ of the time to $W^\pm$ plus massless neutralino LSP the bound is $m_{\chi^{\pm}} > 410$ GeV. The bound gets weaker when the LSP increases in mass since the decay products of the chargino get softer.   As  ATLAS has updated the analysis with the full data set  we will base our discussion on that  analysis, although our conclusions will apply to any search with similar requirements. 

The ATLAS search requires exactly two opposite sign leptons $(e/\mu)$ with $p_T > 25$ GeV and $|\eta|<2.47\, (2.7)$ electrons (muons), and an invariant mass of the dilepton pair greater than $100$ GeV. In addition, the missing transverse energy must be greater than $110$ GeV. Up to one light flavor jet satisfying $p_T>20$ GeV and $|\eta|<2.4$ is allowed, while all events containing a b-tagged jet are vetoed.   Surviving events are further classified according to whether the leptons have the same or different flavor and the number of light flavor (non-$b$ quark or gluon) jets (0 or 1), then broken into several signal regions according to the kinematical variable $M_{T2}=\min_{{\slashed p_{T1}+\slashed p_{T2}=\slashed p_T}}\{\max[m_T(p^{\ell_1}_T, \slashed p_{T1}),m_T(p^{\ell_2}_T, \slashed p_{T2})]\}$, where $p^{\ell_{1,2}}_T$ are the transverse momenta of the leptons and $\slashed p_T$ is the missing transversed momentum. Not seeing any excesses from the SM background ATLAS derives the bound of $410$ GeV for an isolated wino-like chargino and a massless LSP.

 We have shown in the previous section that whenever the chargino is the LSP there are other states close by in mass that will populate the same signal so one has to reinterpret the previous bound in a more realistic situation. To study this in Monte Carlo, we use the MSSM model from Sec.~\ref{sec: models} as a test case, working with a UFO model file that includes the Feynman rules of the MSSM in GMSB~\cite{Christensen:2013aua} (within the framework of \texttt{MadGraph5}\_\texttt{aMC}@\texttt{NLC}~\cite{Alwall:2014hca}). We set the gravitino as the LSP, decouple all sparticles other than the electroweakinos, and (following Sec.~\ref{sec: models}) choose electroweakino masses such that the Higgsino is the lightest multiplet with the chargino lighter than the lightest neutralino. 
 
For a given chargino mass and fixed chargino-neutralino mass difference of $\Delta m = 10\, \gev$, we simulate the production of all possible pairs of chargino/neutralino, $p p\to \chi_i \chi_j$ (where $i,j=1,2,\pm$), forcing the neutralinos $\chi_{1,2}$  to beta decay so that every event contains $W^+ W^-$ + $\slashed E_T +$ soft particles. The parton level events are then passed through \texttt{Phythia8}~\cite{Sjostrand:2014zea} for the $W^{\pm}$ decays, showering and hadronization, then through \texttt{Delphes}~\cite{deFavereau:2013fsa} for detector simulation. We generate $50000$ events for every electroweakino mode  ($\chi^+\chi^-$, $\chi_1\chi_2$, etc.).
 
The simulated events are then run through the ATLAS analysis~\cite{ATLAS:2019cfv} and separated into signal regions.  We find that the total signal efficiency (summing over all signal regions) is between $1.4\%$ and $1.5\%$ for every electroweakino mode -- i.e. $pp \to \chi^0_1\chi^0_2$, $pp \to \chi^0_1\chi^+$, etc. have the same analysis efficiency as $pp \to \chi^+\chi^-$. In addition, there are no appreciable differences in how different electroweakino modes populate the individual signal regions. 

To determine the mass bound in our setup, we equate the  cross section times efficiency for $pp \to \chi_i \chi_j$ ($i,j = 1,2, \pm$, Higgsino-like hierarchy) to the cross section times efficiency for the model ATLAS uses, $pp \to \chi^+\chi^-$ (wino-like). Not knowing the full details of how the different signal regions are combined and weighted in the statistical analysis, we use the total efficiency (summing all signal regions) to set bounds. The LSP is massless in our scenario\footnote{The three models have a massless LSP (or $\phi$ in the non-susy case).}, therefore the ATLAS number we want to compare to is $m_\chi = 410\, \gev$. Because the analysis efficiency is the same for all production modes, it drops out of the equation and the bound is determined by cross sections alone:
\begin{equation}
\sum_{i,j} \sigma(pp \to \chi_i\chi_j)(m_\chi) = \sigma(pp \to \chi^+\chi^-)_{\text{wino}}(m_\chi = 410\,\gev).
\end{equation}
Here, the $m_\chi$ is included to remind us that it's the only parameter we dial (the mass splitting is fixed to $10\,\gev$, all branching fractions are $\sim 100\%$, and the only couplings involved are electroweak gauge couplings). Using the NLO-NNL cross sections from Ref.~\cite{Fuks:2013vua} the cross section for the ATLAS model is $48\, \text{fb}$, which translates into an exclusion bound (95\% CL) on the Higgsino mass of $460\,\gev$. While we have calculated this bound using the MSSM model, it applies to the other scenarios we have presented since the three of them have an electroweak doublet decaying to two leptons plus missing energy (and soft objects).
  
Although we have calculated our bound using the dilepton signal, there are events with three leptons, with the third one coming from the leptonic beta decay of the neutralino. These leptons are too soft to be used to put a bound using the trilepton signal~\cite{Sirunyan:2017qaj,Aaboud:2018sua}, as we have emphasized, but they are still present and are a potential handle to improve the search (or to dig out what model is causing a signal, should one be seen). Specifically, we could look for the presence of a third lepton off-line, where $p_T$ thresholds are typically lower.  Finding a third lepton will indicate that there are several states with similar mass and not just an isolated charged state. In Fig.~\ref{pt} below, we have plotted the $p_T$ of the third lepton for events that have passed all other $W^+W^- + \slashed E_T$ cuts (originating from a spectrum with $\Delta m = 10\, \gev)$. There are about 5 events with $p_T>5$ GeV per 100 fb$^{-1}$, while only $\sim 2$ events above $10\, \gev$. This is an idealized plot, achieved by fixing the lepton id for leptons with $2\, \gev < p_T < 10\,\gev$ to $100\%$ in the \texttt{Delphes} card (rather than the conservative default of $0\%$ for leptons with $p_T < 10\,\gev$), but it does give some idea of what sort of spectrum to expect and how the yield will depend on $p_T$. For smaller $\Delta m$ the $p_T$ of the third lepton will be softer, making them more difficult to reconstruct. 

\begin{figure}[h!]
\centering 
\includegraphics[scale=0.45]{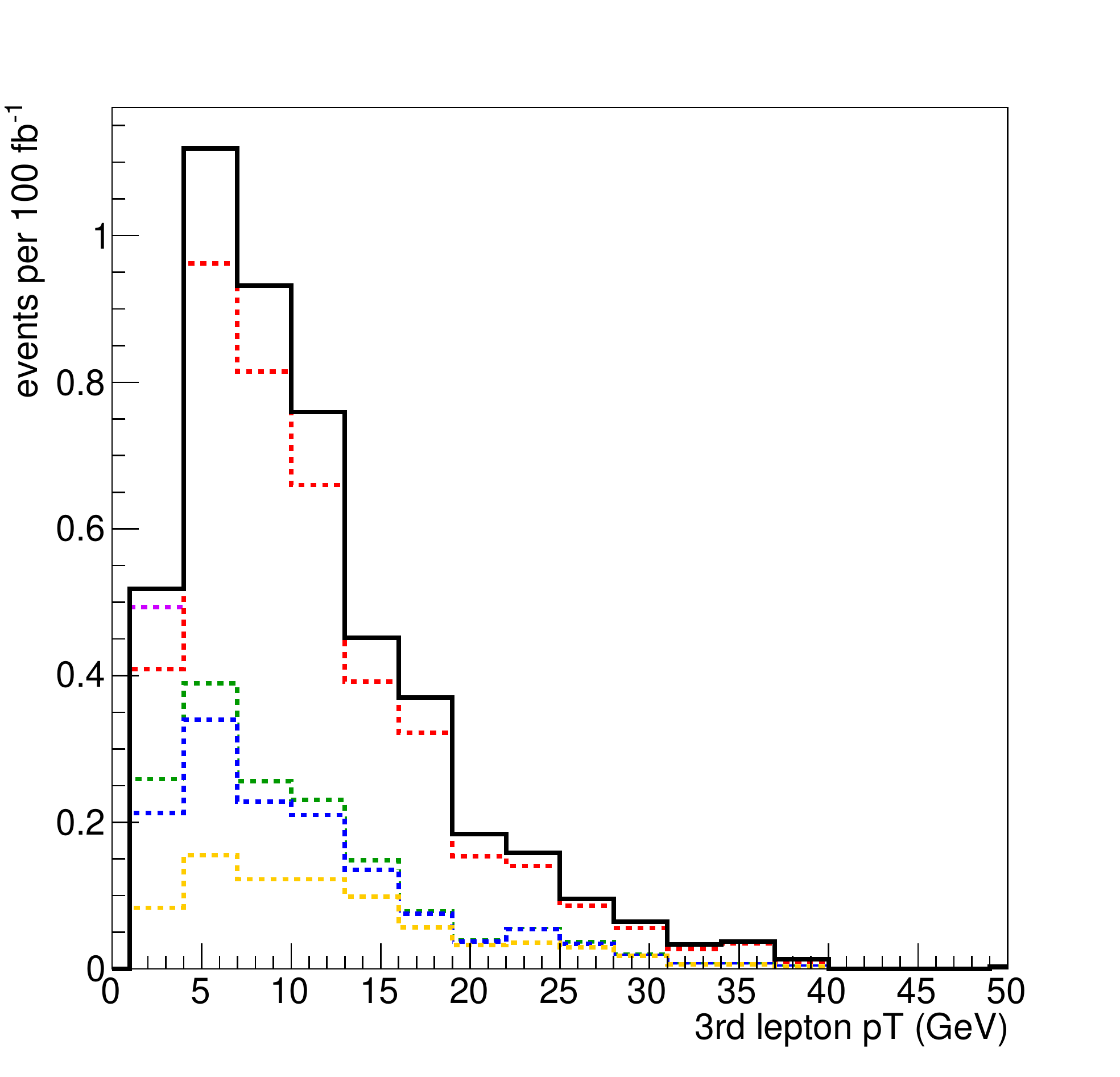}
\caption{The $p_T$ of the third hardest lepton in simulated events that pass the ATLAS analysis. The dotted lines indicate the contributions from (starting from the bottom) $pp \to \chi^0_1\chi^0_2$ (yellow), $\chi^0_1\chi^+$ (blue), $\chi^0_1\chi^-$ (green), $\chi^0_2\chi^+$ (red), $\chi^0_2\chi^-$ (violet), and $\chi^-\chi^+$ (cyan). The black line shows the sum of all contributions.}
\label{pt}
\end{figure}

\section{Conclusions}
\label{sec.conclusions}

 In this paper, we have investigated the circumstances under which electroweakinos in natural supersymmetry escape bounds coming from the $W^\pm Z^0$+$\slashed E_T$ `trilepton' channel.  We identify three criteria: 1.) a compressed electroweakino spectrum, with a chargino as the lightest state and mass splittings to heavier neutralino state(s) $O (10\,\gev)$, 2.) the predominant neutralino decay mode is beta decay, to $\chi^{\pm}+W^{\mp}$, and 3.) a gravitino LSP (or other, neutral, non-electroweakino state). One unavoidable consequence of these criteria is that, all $\chi_i \chi_j$, $i = 1,2,\pm$ modes lead to a $W^{\pm}W^{\mp} + \slashed E_T$ final state (plus additional, soft particles) and must be considered when interpreting experimental limits in that channel.

We provided three example models, two supersymmetric and one non, that realizes the above features. Then, using a MSSM GMSB model with Higgsinos as the lightest electroweakinos, we recast the ATLAS $W^+W^- + \slashed E_T$ analysis, including all electroweakino modes. For mass splittings among all  electroweakinos $\Delta m < 10\, \gev$, we find all electroweakino modes have the same analysis efficiency. The resulting exclusion bound is $m_\chi > 460\, \gev$, compared to the ATLAS bound of $410\,\gev$ (massless LSP).

In general, there are soft leptons coming from the (beta-)decay of the neutralino to the chargino that can potentially be used to distinguish between a model with an isolated charged state from a model with a doublet almost degenerated in mass.

\section*{Acknowledgments}

This work was partially supported by the National Science Foundation under grant PHY-1820860. We would like to thank Benjamin Fuks for helping us with the implementation of the Feynrules model and Zach Marshall for communication about the ATLAS search.

\appendix
\section{R-symmetric superpotential}
\label{app:mrssm}
 In the R-symmetric MSSM (MRSSM)~\cite{Kribs:2007ac, Dudas:2013gga, Diessner:2014ksa} the $U(1)_R$ symmetry inherent in supersymmetric kinetic terms is imposed on the superpotential. The $U(1)_R$ charges of the gauge fields (and their superpartners) are fixed $= 1$, but there is some flexibility in the matter sector. In order for EWSB to not spoil $U(1)_R$, we require $R_{H_u} = R_{H_d} = 0$. All other MSSM matter fields are given $R = 1$. As the superpotential must have $R = 2$, this charge assignment forbids the usual $\mu$-term. Without this term, Higgsinos would be massless. To fix this issue, we introduce new superfields, R-Higgses, with the same SM quantum numbers as the Higgs but carrying $R$-charge $=2$. We can then write gauge and $U(1)_R$ invariant mass terms connecting the MSSM Higgses to their R-partners. As there are two MSSM Higgses, we need two R-Higgses:
\begin{align}
\mathcal W \supset \mu_u\, R_u H_u + \mu_d\, R_d\ H_d
\end{align}
We assume the R-Higgses do not get vevs. Having added the R-Higgses, we need to assess whether there are other interactions we need to include. Trilinear interactions involving R-Higgses and two MSSM fields are forbidden by R-symmetry, but we can write down superpotential trilinear interactions between R-Higgses, MSSM Higgses, and the $SU(2)_w$ and $U(1)_B$ Dirac mass partners $\Phi_{W_a}, \Phi_B$ (which carry R-charge $= -1$):
\begin{align}
\mathcal W \supset \lambda_u\, R_u \tau^a H_u \Phi_{W_a} + \lambda_d\, R_d\ \tau^a\, H_d\, \Phi_{W_a} +  \lambda'_u\, R_u H_u \Phi_{B} + \lambda'_d\, R_d H_d\, \Phi_{B},
\end{align}
where $\tau^a$ are $SU(2)_w$ generators. Once EWSB occurs, these $\lambda$ interactions lead to mixing among the fermionic components in $H_u, R_u, \Phi_W, \Phi_B$, the full set of MRSSM electroweakinos. In section \ref{sec: models}, we make the simplifying assumption that $\lambda_{u,d} \sim g, \lambda'_{u,d} \sim g'$, and $\mu_u = \mu_d$.

\end{document}